\documentstyle[12pt,epsf]{article}
\makeatletter
\@addtoreset{equation}{section}

\makeatother

\begin{document}

\begin{flushright}
Dec 2000

OU-HET 375
\end{flushright}

\begin{center}

\vspace{5cm}
{\Large BPS Vortices in Brane-Antibrane Effective Theory} 

\vspace{2cm}
Takao Suyama \footnote{e-mail address : suyama@funpth.phys.sci.osaka-u.ac.jp}

\vspace{1cm}

{\it Department of Physics, Graduate School of Science, Osaka University, }

{\it Toyonaka, Osaka, 560-0043, Japan}

\vspace{4cm}

{\bf Abstract} 

\end{center}

We investigate an action which resembles the effective action of brane-antibrane system 
derived from boundary string field theory. 
We find that the action has smooth vortex solutions which saturate 
the Bogomol'nyi bound. 

\newpage

\vspace{1cm}

\section{Introduction}

\vspace{5mm}

The study of tachyon condensation \cite{TC} provides us many insights into the off-shell 
dynamics of string theory. 
It has been established that open string tachyon is stabilized by acquiring a nonzero vev. 
As a result of this condensation, an unstable D-brane system decays into the vacuum. 
The occurrence of this phenomenon is shown by calculating the tachyon potential by using 
string field theory (SFT) \cite{V(T)} or boundary string field theory (B-SFT) \cite{B-SFT}. 
In addition, the notion of the D-brane charge is refined by regarding the D-brane as a 
remnant of the decay. 
The study of the unstable D-brane system leads us to introduce K-theory into string theory 
\cite{Ktheory}. 

Suppose that the effective field theory of the unstable D-brane system is known. 
Then the action should have nontrivial solutions which describe a lower dimensional D-brane 
system. 
So it is expected that the effective theory can be used to describe various D-brane systems 
as classical solutions of the theory. 

The unstable D-brane system can also decay into a BPS D-brane system. 
For example, Dp-$\bar{\mbox{D}}$p system can decay into a D(p$-$2)-brane. 
Therefore the effective theory of the brane-antibrane system should have a BPS vortex 
solution corresponding to the D(p$-$2)-brane. 
So the search for such solution is an interesting check for the validity of the action of 
the effective theory. 

Recently, the effective action of the brane-antibrane system is derived by using B-SFT 
\cite{DDbar}. 
The leading order terms in the derivative expansion are the followings,
\begin{eqnarray}
S_{p-\bar{p}} &=& 2\tau_9\int d^{10}x\ e^{-2\pi\alpha'|T|^2}\left[\ 
                  1+8\pi\alpha'\log2|D_\mu T|^2 \right. \nonumber \\
              & &\left. +\frac{(2\pi\alpha')^2}8(F_{\mu\nu}^+)^2
                 +\frac{(2\pi\alpha')^2}8(F_{\mu\nu}^-)^2
                 +\frac\beta8(\alpha')^2|T|^2(F_{\mu\nu}^+-F_{\mu\nu}^-)^2 \ \right],
  \label{BSFT}
\end{eqnarray}
where $F_{\mu\nu}^+$ is the field strength of the gauge field $A_\mu^+$ on the D-brane and 
$F_{\mu\nu}^-$ is the one on the $\bar{\mbox{D}}$-brane. 
The covariant derivative of the tachyon field $T$ is defined as 
\begin{equation}
D_\mu T = \partial_\mu T-i(A_\mu^+-A_\mu^-)T.
\end{equation}

There is another progress on the configuration of $T$ and $A_\mu^{\pm}$ \cite{holo}. 
It is shown from the worldsheet analysis that $T$ must be holomorphic on the associated line 
bundle in order to preserve worldsheet supersymmetry, i.e.
\begin{equation}
D_{\bar{a}}T=0,
  \label{holo}
\end{equation}
where $a$ is an index for the complex coordinates. 
So this implies that one of the vortex equation is (\ref{holo}). 

In this paper we will investigate an action which resembles the above action (\ref{BSFT}) 
and find a set of BPS solutions with smooth behavior, which corresponds to 
codimension two D-branes. 
It seems natural to identify these solutions with BPS D-branes since they saturate the 
Bogomol'nyi bound. 
It should be noted that these solutions are, in fact, not the classical solutions of the 
action (\ref{BSFT}). 
They are the solutions of the action with slightly different potential. 
This can be explained as follows. 
The exact vortex solution would depend on higher derivative terms which we will omit 
throughout this paper. 
So the action (\ref{BSFT}) not necessarily has BPS vortex solutions. 
It is nonetheless interesting that an action with nontrivial metric in kinetic terms and a 
run-away potential has regular BPS vortex solutions. 

This paper is organized as follows. 
In section 2 we will review our previous work on a description of BPS D-brane systems 
\cite{TSonDDbar}. 
We will discuss an action which resembles the action (\ref{BSFT}) in section 3 and find a 
set of first order equations. 
The properties of the solution are investigated in section 4.

\vspace{1cm}

\section{BPS D-branes as vortices}

\vspace{5mm}

In this section, we will review our previous work \cite{TSonDDbar} and show a description of 
BPS D-brane systems. 
We consider the effective field theory of D9-$\bar{\mbox{D}}$9 system. 
The bosonic part consisting of terms which are leading order in $\alpha'$ is the following 
\cite{leading},
\begin{equation}
S_{boson} = \frac1{g_{YM}^2}\int d^{10}x\ \left[\ 
           -\frac14F_{\mu\nu}F^{\mu\nu}-|D_\mu T|^2-\frac12(|T|^2-\zeta)^2\ \right],
  \label{U(1)}
\end{equation}
where $D_\mu T=\partial_\mu T-iA_\mu T$. 
The gauge field $A_\mu$ corresponds to a linear combination $A_\mu^+-A_\mu^-$ of the gauge 
fields in the action (\ref{BSFT}). 
We will find BPS solutions of this action and show the correspondence to the BPS D-brane 
systems. 

It should be noted that when one considers the leading terms in the derivative expansion, 
each kinetic term would have a nontrivial metric. 
These metrics are important when one discusses the dynamics around the vacuum configuration 
\cite{vacuum}. 
On the other hand, for the discussion of the D-brane these metrics are not so important 
since the worldvolume of the D-brane corresponds to the zero set of $T$. 
Around $T=0$ the metrics do not vanish and can be approximated by constants. 
So we expect that the worldvolume of the D-brane system can be described, at least 
qualitatively, by a solution of the action (\ref{U(1)}). 

The action (\ref{U(1)}) has BPS solutions satisfying the following equations. 
\begin{eqnarray}
&& D_{\bar{a}}T=0, \hspace{1cm} F_{\bar{a}\bar{b}}=0 \label{BPS1} \\
&& -ig^{a\bar{b}}F_{a\bar{b}}+|T|^2-\zeta=0 \label{BPS2}
\end{eqnarray}
We have introduced the complex coordinates
\begin{equation}
z^a=x^{2a-1}+ix^{2a}
\end{equation}
and $a,b=1,\cdots,n$. 
The case $n=1$ corresponds to the Nielsen-Olesen vortex equations \cite{vortexNO}. 
The cases $n=2,3$ describe intersecting branes each of which has codimension two. 

The BPS equations (\ref{BPS1})(\ref{BPS2}) are well-studied when they are defined on a 
compact K\"ahler manifold $M$ \cite{tauvortex}. 
There is one-to-one correspondence between the set of solutions of the BPS equations 
and the set of holomorphic sections $T$ of the line bundles on $M$. 
The zero set of $T$ is an effective divisor of $M$, so the solutions describe various BPS 
D-brane systems wrapped on the divisor of $M$ \cite{wrappedD}. 
Thus this description can apply to multiple branes and intersecting branes as well as a 
single brane.

\vspace{1cm}

\section{BPS equations}

\vspace{5mm}

In this section we will discuss a more general action.
\begin{equation}
S = -\frac1{g_{YM}^2}\int d^{10}x\ \left[\  e^{-a|T|^2}|D_\mu T|^2
    +\frac14e^{-b|T|^2}(1+c|T|^2)F_{\mu\nu}F^{\mu\nu}+V(T)\  \right]
  \label{action}
\end{equation}
We assume that $a,b,c$ are positive constants. 
As will be shown below, for special form of $V(T)$ this action has BPS solutions which 
satisfy a set of first order equations. 

We will look for the solutions which depends only on two spatial coordinates, say, 
$x^1$ and $x^2$. 
The energy density of the solution is as follows, 
\begin{equation}
{\cal E} = \frac1{g_{YM}^2}\int d^2x\ \left[\  e^{-a|T|^2}|D_\alpha T|^2
          +\frac14e^{-b|T|^2}(1+c|T|^2)F_{\alpha\beta}F_{\alpha\beta}+V(T) \ \right],
   \label{density}
\end{equation}
where $\alpha,\beta=1,2$. 

The following identity holds. 
\begin{eqnarray}
e^{-a|T|^2}|D_\alpha T|^2 &=& e^{-a|T|^2}|D_1T+iD_2T|^2
                             -\frac1a\left(e^{-a|T|^2}-1\right)F_{12}
                            \nonumber \\
                          & &+\frac ia\epsilon_{\alpha\beta}\partial_\alpha\left\{
                              (e^{-a|T|^2}-1)\frac{D_\beta T}T\right\}
  \label{identity}
\end{eqnarray}
This identity is proved in Appendix \ref{AppA}. 
Therefore the energy density (\ref{density}) can be rewritten as 
\begin{eqnarray}
{\cal E} &=& \frac1{g_{YM}^2}\int d^2x\ \left[\  e^{-a|T|^2}|D_1T+D_2T|^2
            +\frac12e^{-b|T|^2}(1+c|T|^2)
            \left(F_{12}-\frac1a\frac{e^{-(a-b)|T|^2}}{1+c|T|^2}\right)^2
            \right.  \nonumber \\
         & &\hspace{1cm} +\frac1aF_{12}+\frac ia\epsilon_{\alpha\beta}\partial_\alpha\left\{
             (e^{-a|T|^2}-1)\frac{D_\beta T}T\right\} 
            \left. +V(T)-\frac1{2a^2}\frac{e^{-(2a-b)|T|^2}}{1+c|T|^2} 
            \ \right]. \nonumber \\
\end{eqnarray}
So when the potential takes the following form 
\begin{equation}
V(T) = \frac1{2a^2}\frac{e^{-(2a-b)|T|^2}}{1+c|T|^2} ,
\end{equation}
one can find a set of first order equations 
\begin{eqnarray}
&& D_1T+iD_2T = 0, \\
&& F_{12}-\frac1a\frac{e^{-(a-b)|T|^2}}{1+c|T|^2}=0, 
\end{eqnarray}
whose solutions saturates the Bogomol'nyi bound. 
As will be shown in the next section, the energy density of the BPS solution is quantized.  

An interesting case in relation to the brane-antibrane system is the case $a=b$. 
In this case the action (\ref{action}) resembles the one derived from the B-SFT 
\cite{B-SFT}, 
and has BPS solutions which satisfy
\begin{eqnarray}
&& D_1T+iD_2T = 0, \label{vortex1} \\
&& F_{12}-\frac1{a(1+c|T|^2)} = 0. \label{vortex2}
\end{eqnarray}

\vspace{1cm}

\section{Properties of the solutions}

\vspace{5mm}

We will focus on the case $a=b$. 
Equation (\ref{vortex1}) can be solved by parametrizing $T$ and $A_\alpha$ as follows. 
\begin{eqnarray}
T &=& \rho^{\frac12}e^{i\omega} \\
A_1 &=& \frac12\partial_2\log\rho+\partial_1\omega \\
A_2 &=& -\frac12\partial_1\log\rho+\partial_2\omega
\end{eqnarray}
Another equation (\ref{vortex2}) is rewritten as
\begin{equation}
-\frac12\Delta\log\rho = \frac1{a(1+c\rho)}.
  \label{eq1}
\end{equation}
We will take the polar coordinates $(r,\theta)$ and look for radially symmetric solutions. 
If $\rho$ does not depend on the angle variable $\theta$, equation (\ref{eq1}) is 
\begin{equation}
-\frac12\left[ \frac{d^2}{dr^2}+\frac1r\frac d{dr}\right]\log\rho = \frac1{a(1+c\rho)}. 
  \label{eq2}
\end{equation}

It is convenient to make a change of variables
\begin{equation}
u=\log\rho, \hspace{1cm} t=\log r.
\end{equation}
Then equation (\ref{eq2}) is a bit simplified. 
\begin{equation}
\frac{d^2u}{dt^2} = -\frac2a\frac{e^{-2t}}{1+ce^u}
  \label{eq3}
\end{equation}

We have to specify the boundary condition of the solution to solve the equation (\ref{eq3}). 
From the form of the potential, one can see that the minima of the potential correspond to 
$|T|\to\infty$. 
So we have to impose $u\to+\infty$ at infinity $(t\to+\infty)$ for a localized solution. 
On the other hand, the core of the solution $(r=0)$ should correspond to $|T|=0$, so 
$u\to -\infty$ as $t\to -\infty$. 

The asymptotic behavior of the solution can be derived from equation (\ref{eq3}),
\begin{equation}
u \sim \left\{
  \begin{array}{lc}
     2At & (t\to +\infty), \\ 2nt & (t\to -\infty),
  \end{array}
\right.
\end{equation}
where $A\ge1$ and $n$ is a positive integer.  
So there exist solutions which satisfy the desired boundary condition. 
The form of a solution is shown in fig.1,2. 
For details see Appendix \ref{AppB}. 

There is one exceptional case; $n=1$. 
Since $\frac{d^2u}{dt^2}<0$ everywhere, the coefficient $A$ is restricted so that $A<n$. 
So in the case $n=1$ there is no solution with the desired boundary behavior. 

\vspace{2mm}

The energy density of the solution is 
\begin{equation}
{\cal E} = \frac1{g_{YM}^2a}\int d^2x\ \left[\ F_{12}
   +i\epsilon_{\alpha\beta}\partial_\alpha\left\{(e^{-a|T|^2}-1)\frac{D_\beta T}T
   \right\}\ \right].
  \label{energy}
\end{equation}
Since $|T|\to \infty$ at infinity, ${\cal E}$ is finite. 
In fact, 
\begin{eqnarray}
{\cal E} &=& \frac1{g_{YM}^2a}\lim_{R\to\infty}\int_{S^1_R}dx^\alpha 
             \left[\ A_\alpha+i(e^{-a|T|^2}-1)\frac{D_\alpha T}T\ \right] \nonumber \\
         &=& \frac1{g_{YM}^2a}\lim_{R\to\infty}\int_{S^1_R}dx^\alpha 
             \left[\ A_\alpha-i\frac{D_\alpha T}T\ \right] \nonumber \\
         &=& \frac1{g_{YM}^2a}\lim_{R\to\infty}\int_{S^1_R}dx^\alpha 
             \left[\ -\frac i2\partial_\alpha\log\rho+\partial_\alpha\omega\ \right]
             \nonumber \\
         &=& \frac1{g_{YM}^2a}\left[\ 2\pi n
            -iA\lim_{R\to\infty}\int_{S^1_R}dx^\alpha \frac{x^\alpha}{r^2}\ \right]
             \nonumber \\
         &=& \frac{2\pi}{g_{YM}^2a}n,
\end{eqnarray}
where $S^1_R$ denotes a circle with radius $R$. 
Therefore ${\cal E}$ is positive and quantized. 
So the BPS solutions are stable which is appropriate for the correspondence to the BPS 
D-branes. 

\vspace{2mm}

As shown above, there are stable solutions labelled by $n$. 
However the flux is not quantized and depends on the behavior of $T$ at infinity. 
So it is hard to interpret the flux as the number of codimension two branes. 
It would be rather appropriate for equations (\ref{vortex1})(\ref{vortex2}) to be 
defined on a compact surface. 
Then the flux is automatically quantized. 
Since the second term in (\ref{energy}) vanishes, the energy density is still quantized 
and the solutions are stable. 
It is very interesting to show the existence of the solutions on the compact surface. 
 
\vspace{1cm}

{\Large {\bf Acknowledgements}}

\vspace{5mm}

I would like to thank H. Itoyama, H. Kawai and T. Matsuo for valuable discussions. 
This work is supported in part by JSPS Reseach Fellowships.

\vspace{3cm}

\begin{figure}[htb]
 \epsfxsize=20em
 \centerline{\epsfbox{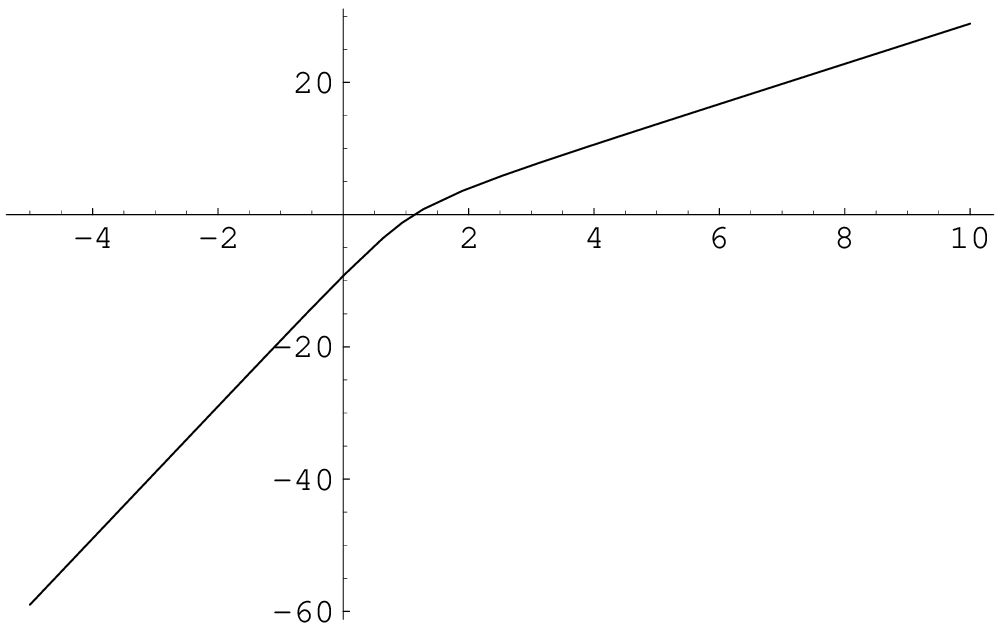}}
 \caption{$u$ vs. $t$ for the case $n=10$.}
\vspace{5mm}
 \epsfxsize=20em
 \centerline{\epsfbox{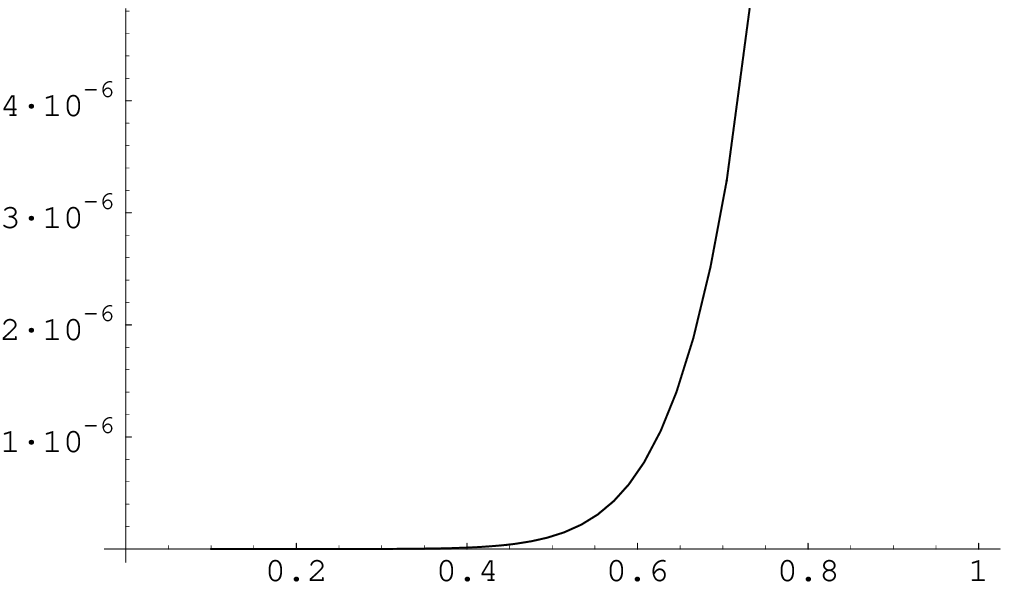}}
 \caption{$\rho$ vs. $r$.}
\end{figure}

\newpage

\appendix

{\LARGE {\bf Appendices}}

\vspace{5mm}

\section{Proof of the identity} \label{AppA}

\vspace{5mm}

One can easily see that 
\begin{equation}
|T|^{2k}|D_\alpha T|^2 = |T|^{2k}|D_1T+D_2T|^2
           -i|T|^{2k}\epsilon_{\alpha\beta}D_\alpha T^\dagger D_\beta T.
\end{equation}
The second term in the RHS is rewritten as, 
\begin{eqnarray}
-i|T|^{2k}\epsilon_{\alpha\beta}D_\alpha T^\dagger D_\beta T
&=& ik|T|^{2k}\epsilon_{\alpha\beta}D_\alpha T^\dagger D_\beta T
   +\frac12|T|^{2k}\epsilon_{\alpha\beta}|T|^2F_{\alpha\beta} \nonumber \\
& &-i\partial_\alpha\left(|T|^{2k}\epsilon_{\alpha\beta}T^\dagger D_\beta T\right).
\end{eqnarray}
Thus 
\begin{eqnarray}
-i|T|^{2k}\epsilon_{\alpha\beta}D_\alpha T^\dagger D_\beta T
&=& \frac1{2(k+1)}|T|^{2(k+1)}\epsilon_{\alpha\beta}F_{\alpha\beta} \nonumber \\
& &-i\partial_\alpha\left(\frac1{k+1}|T|^{2k}\epsilon_{\alpha\beta}T^\dagger D_\beta T
    \right).
\end{eqnarray}
This identity can be used to prove the identity (\ref{identity}). 
\begin{eqnarray}
e^{-a|T|^2}|D_\alpha T|^2
&=& e^{-a|T|^2}|D_1T+iD_2T|^2
   -i\sum_{k=0}^\infty \frac{(-a)^k}{k!}|T|^{2k}
   \epsilon_{\alpha\beta}D_\alpha T^\dagger D_\beta T \nonumber \\
&=& e^{-a|T|^2}|D_1T+iD_2T|^2
   +\frac12\sum_{k=0}^\infty \frac{(-a)^k}{(k+1)!}|T|^{2(k+1)}
   \epsilon_{\alpha\beta}F_{\alpha\beta} \nonumber \\
& &-i\partial_\alpha\left(\sum_{k=0}^\infty\frac{(-a)^k}{(k+1)!}|T|^{2k}
    \epsilon_{\alpha\beta}T^\dagger D_\beta T\right) \nonumber \\
&=& e^{-a|T|^2}|D_1T+iD_2T|^2
   -\frac1{2a}\left(e^{-a|T|^2}-1\right)\epsilon_{\alpha\beta}F_{\alpha\beta} \nonumber \\
& &+\frac ia\epsilon_{\alpha\beta}\partial_\alpha\left\{(e^{-a|T|^2}-1)\frac{D_\beta T}T
    \right\}
\end{eqnarray}

\vspace{1cm}

\section{Asymptotic behavior of the solutions} \label{AppB}

\vspace{5mm}

Solving equations (\ref{vortex1})(\ref{vortex2}) is reduced to solving the equation 
\begin{equation}
\frac{d^2u}{dt^2} = -\frac2a\frac{e^{2t}}{1+ce^u}.
  \label{equationofu}
\end{equation}

First consider the region $u\to -\infty$. 
Then $e^u$ in the RHS can be neglected and the solution behaves as
\begin{equation}
u \sim -\frac1{2a}e^{2t}+Ct+D.
\end{equation}
So $u$ is linear in $t$ at $t\to -\infty$. 
This means that $\rho$ behaves as 
\begin{equation}
\rho \sim r^C
\end{equation}
near the origin. 
The regularity and the single-valuedness of the solution require that 
\begin{equation}
\omega = n\theta, \hspace{1cm} C=2n.
\end{equation}

Next consider the region $u\to +\infty$. 
In this region equation (\ref{equationofu}) is 
\begin{equation}
\frac{d^2u}{dt^2} = -\frac2{ac}e^{2t-u}.
\end{equation}
One can easily integrate this equation and obtains
\begin{equation}
\left(\frac{dv}{dt}\right)^2=\frac4{ac}e^{-v}+\xi,
\end{equation}
where $v=u-2t$ and $\xi$ is an integration constant. 
This equation can be solved exactly,
\begin{equation}
e^{-u+2t} = \left\{
  \begin{array}{lc}
     ac\xi e^{-\sqrt{\xi}(t-t_0)}(1-e^{-\sqrt{\xi}(t-t_0)})^{-2}, & 
     (\xi>0) \\
     ac(t-t_0)^{-2}, & (\xi=0) \\
     -\frac{\xi ac}4\sec^2\frac{\sqrt{|\xi|}}2(t-t_0), & (\xi<0)
  \end{array}
\right.
\end{equation}
where $t_0$ is another integration constant. 
The solution for $\xi<0$ is singular and this is not appropriate for the asymptotic 
solution. 
The other solutions behave at $t\to+\infty$ as
\begin{equation}
u \sim (2+\sqrt{\xi})t.
\end{equation}
Therefore $u$ is also linear in $t$ at $t\to+\infty$ although its slope is restricted. 

\newpage

\end{document}